\documentclass[fleqn,usenatbib]{mnras}

\usepackage{newtxtext,newtxmath}
\usepackage[T1]{fontenc}
\usepackage{ae,aecompl}
\usepackage{array}
\usepackage{tabu}

\usepackage{graphicx}	% Including figure files
\usepackage{amsmath}	% Advanced maths commands
\usepackage{amssymb}	% Extra maths symbols
\newcommand{\HI}{H\,{\sc i}}

\title[\HI\ gas in Jellyfish Galaxy JO204]{GASP XXV: Neutral Hydrogen gas in the striking Jellyfish Galaxy JO204}

\author[T. Deb et al.]{
Tirna Deb,$^{1}$\thanks{E-mail: deb@astro.rug.nl (TD)}
Marc A. W. Verheijen$^{1,2}$,
Marco Gullieuszik$^{3}$,
Bianca M. Poggianti$^{3}$,
\newauthor
Jacqueline H. van Gorkom$^{4}$,
Mpati Ramatsoku$^{5}$,
Paolo Serra$^{5}$,
Alessia Moretti$^{3}$,
 \newauthor
 Benedetta Vulcani$^{3}$,
 Daniela Bettoni$^{3}$,
 Yara L., Jaff\'{e}$^{6}$,
 Stephanie Tonnesen$^{7}$
 \newauthor
and Jacopo Fritz$^{8}$\\ \\\\
$^{1}$Kapteyn Astronomical Institute, University of Groningen, Postbus 800, NL-9700 AV Groningen, The Netherlands\\
$^{2}$National Centre for Radio Astrophysics, Tata Institute of Fundamental Research,  Postbag 3, Ganeshkhind, Pune 411 007, India\\
$^{3}$INAF- Osservatorio Astronomico di Padova, Vicolo dell'Osservatorio 5, I-35122 Padova, Italy\\
$^{4}$Department of Astronomy, Columbia University, Mail Code 5246, 550 W 120th Street, New York, NY 10027, USA\\
$^{5}$INAF- Osservatorio Astronomico di Cagliari, Via della Scienza 5, I-09047 Selargius (CA), Italy\\
$^{6}$Instituto de F\'{i}sica y Astronom\'{i}a, Facultad de Ciencias, Universidad de Valpara\'{i}so, Avda. Gran Bretana 1111 Valpara\'{i}so, Chile\\
$^{7}$Center for Computational Astrophysics, Flatiron Institute, 162 5th Ave, New York, NY 10010, USA\\
$^{8}$Instituto de Radioastronomia y Astrofisica, UNAM, Campus Morelia, A.P. 3-72, C.P. 58089, Mexico\\
}

\date{Accepted 2020 April 1. Received 2020 March 27; in original form 2019 September 23}

\pubyear{2019}

\begin{document}
\label{firstpage}
\pagerange{\pageref{firstpage}--\pageref{lastpage}}
\maketitle

\begin{abstract}

We present JVLA-C observations of the \HI\ gas in JO204, one of the most striking jellyfish galaxies from the GASP survey. JO204 is a massive galaxy in the low-mass cluster Abell 957 at z=0.04243. The \HI\ map reveals an extended 90 kpc long ram-pressure stripped tail of neutral gas, stretching beyond the 30 kpc long ionized gas tail and pointing away from the cluster center. The \HI\ mass seen in emission is (1.32 $ \pm 0.13) \times 10^{9} \rm M_{\odot}$, mostly located in the tail. The northern part of the galaxy disk has retained some \HI\ gas, while the southern part has already been completely stripped and displaced into an extended unilateral tail. Comparing the distribution and kinematics of the neutral and ionized gas in the tail indicates a highly turbulent medium. Moreover, we observe associated \HI\ absorption against the 11 mJy central radio continuum source with an estimated \HI\ absorption column density of 3.2 $\times 10^{20}$ cm$^{-2}$. The absorption profile is significantly asymmetric with a wing towards higher velocities. We modelled the \HI\ absorption by assuming that the \HI\ and ionized gas disks have the same kinematics in front of the central continuum source, and deduced a wider absorption profile than observed. The observed asymmetric absorption profile can therefore be explained by a clumpy, rotating \HI\ gas disk seen partially in front of the central continuum source, or by ram-pressure pushing the neutral gas towards the center of the continuum source, triggering the AGN activity.

\end{abstract}


%%%%%%%%%%%%%%%%%%%%%%%%%%%%%%%%%%%%%%%%%%%%%%%%%%

%%%%%%%%%%%%%%%%% BODY OF PAPER %%%%%%%%%%%%%%%%%%

\section{Introduction}

\par Galaxies reside in different cosmic environments, from sparsely populated voids to large-scale filaments and densely populated clusters. Depending on the galaxy density of the various environments, galaxies have different star formation and evolutionary histories and display different morphologies. Star-formation activity in galaxies had a peak at z$\approx$2 and has strongly declined afterwards \citep{madau2014,vanderWel2014}. Since then, many star forming galaxies have evolved into `quenched' or passive systems while the star formation rate for a fixed stellar mass has also reduced over time \citep{Noeske2007,Bell2007,Daddi2007,Karim2011}. Along with this evolution of the star-formation activity, a change in galaxy morphologies from `late' to `early' type is also observed in the densest regions \citep{Dressler1997,Fasano2000,Postman2005,Smith2005,Capak2007,Poggianti2009,vulcani2011}.

\par Gas is the fuel for star formation and an effective tracer of environmental effects and internal feedback \citep{Larson1972,Kennicutt1998,Veilleux2005,Silk2012}. To understand the reasons for the decline of star-formation and the origin of various galaxy morphologies, we need to understand the internal and external physical processes of gas accretion and removal from galaxies. Among the internal processes, cold gas can be accreted by the cooling of hot gas in the dark matter halo, that subsequently becomes part of the interstellar medium (ISM) of the galactic disk \citep{White1978,Efstathiou1983,White1991,fraternali2017}. \cite{sancisi2008} studied cold gas accretion from minor mergers and concluded that the accreted gas is insufficient to replenish the gas consumed by star formation.

 On the other hand, the cold ISM may be removed through stellar or AGN feedback, resulting in a decline in star formation.
There are also several external processes that can affect the cold gas content of galaxies, as reviewed in \cite{Boselli2006,boselli2014}. Most of these external gas removal mechanisms have strong environmental dependencies and affect the star formation and evolutionary history of a galaxy -- possibly being responsible for the morphology-density relation \citep{Dressler1980}.

\par There are two main effects of the environment on the evolution of galaxies: gravitational and hydrodynamical. Gravitational interactions affect both the stellar and gaseous components while hydrodynamical interactions only affect the gaseous component. Gravitational perturbations such as tidal galaxy-galaxy interactions and mergers \citep{Spitzer1951,Tinsley1979,Merritt1983,Springel2000}, galaxy-cluster interactions \citep{Byrd1990,Valluri1993}, and galaxy harassments \citep{Moore1996,Jaffe2016} generally happen in denser environments such as galaxy groups and cluster outskirts. During these interactions, the cold gas may also fall towards the central regions and act as fuel for nuclear star-formation or feed a supermassive black hole,triggering an active galactic nucleus \citep[AGN,][]{Baldry2004, Balogh2009}. The processes that do not affect the stellar disk directly are `starvation' \citep{Larson1980,Balogh2000}, thermal evaporation \citep{Cowie1977} and viscous stripping \citep{nulsen1982}.

\par One of the most effective gas removal mechanisms acting on galaxies in dense environments such as clusters is hydrodynamical ram-pressure stripping (RPS). This occurs when galaxies fall into the core of a cluster and encounter a hydrodynamical pressure while passing through the X-ray emitting, hot (T$\sim$ 10$^{7}$-10$^{8}$ K) and dense ($\rho \sim 10^{-3}/cm^{3}$, \citealt{sarazin1986}) intracluster medium (ICM). Ram-pressure $P_{r}$ is proportional to $\rho v^{2}$ where $\rho$ and $v$ stand for the ICM gas density and the speed of the galaxy relative to the ICM, respectively. If the ram-pressure force exceeds the local gravitational restoring force of the galaxy disk, the gas will be stripped out of the gravitational potential \citep{Gunn1972,Faltenbacher2006,Takeda1984,Moran2007,Porter2008}. Extreme examples of RPS are the so called `jellyfish' galaxies that have `tentacles' of material that stretch tens of kpc beyond their disks \citep{Smith2010,Fumagalli2014,Ebeling2014}. The parent sample of GASP (the atlas from which the GASP target galaxies were selected) was based on broad-band B images, and provided a sample of `stripping candidates'. For the purpose of GASP, jellyfish galaxies are defined as the galaxies with a long ionized gas tail (H$\alpha$ tail longer or at least as long as the diameter of the stellar disk). These tentacles show signatures of RPS that creates tail-like structures out of the disk, stimulating star-formation within the tails by collapsing the molecular clouds due to thermal instabilities, turbulent motion etc \citep{poggianti2019}. 

\par The GAs Stripping Phenomena in galaxies (GASP) survey \citep{Poggianti2017} is a European Southern Observatory (ESO) Large Program aimed at observing a statistically significant sample of jellyfish galaxies. Using MUSE integral-field spectrograph on the VLT, 114 galaxies including 64 gas stripping candidates in clusters and 38 in low density environments at z=0.04--0.07 were observed from the \cite{Poggianti2016} sample -- a collection of galaxies from the WINGS \citep{Fasano2006, Moretti2014}, OMEGAWINGS \citep{Gullieuszik2015,Moretti2017} and PM2GC \citep{Calvi2011} samples. The optical images of those galaxies show unilateral debris or disturbed morphologies or multiple star forming knots, indicating a gas-only removal mechanism. They are found in various cosmic environments covering a range of galaxy masses \citep{Poggianti2017}. The key scientific motivation for GASP is to study the  interplay between various gas phases and the star formation activity in different environments.

\par One of the key methods to study gas removal in galaxies is to observe the neutral hydrogen (\HI) gas. Cold \HI\ gas is the primary constituent of the ISM, is generally diffuse and extends well beyond the stellar disk. With the gravitational force being weaker at larger radii, the \HI\ gas can be easily stripped in dense environments such as galaxy groups \citep{Williams1987,Verdes2001,Rasmussen2008,Serra2013} or clusters \citep{Giovanelli1985,Bravo2001,Jaffe2015}. To understand the stripping phenomena thoroughly, neutral hydrogen observations are indispensable for the GASP sample of jellyfish galaxies.

\par Previously, the neutral and ionized gas phases in galaxies experiencing RPS, were studied only in a few nearby clusters. \cite{Oosterloo2005} discovered a 110 kpc long \HI\ plume stripped out of NGC 4388 near the center of the Virgo cluster. In this case, however, the hot halo gas of the M86 group is responsible for the RPS instead of the overall ICM. NGC 4388 also displays a H$\alpha$ tail, which extends only 35 kpc \citep{yoshida2002}, pointing in the same direction. \cite{Yagi2007} found an extremely long and narrow (60 kpc $\times$ 2 kpc) H$\alpha$ tail extending out of the post-starburst galaxy D100 in the Coma cluster for which \cite{bravo-alfaro2000} could not find any \HI\ counterpart. The Virgo cluster galaxy IC 3418 is a ``smoking gun" example of the transformation of a dwarf irregular to a dwarf elliptical galaxy as a result of RPS. In IC 3418, no \HI\ nor H$\alpha$ emission is detected in the main body of the galaxy, but only in the tail \citep{chung2009,kenney2014}. NGC 4569, the brightest late-type galaxy in the Virgo cluster, has a truncated radial \HI\ distribution \citep{chung2009} and long tails of diffuse ionized gas, which suggests that the gas is ionized within the tail during the stripping process \citep{boselli2016}. Interestingly, the Virgo cluster galaxy NGC 4424 in the VESTIGE survey is found to have a H$\alpha$ tail \citep{boselli2018} in the direction opposite to the ram-pressure stripped \HI\ tail \citep{chung2007, chung2009, sorgho2017}. These examples illustrate the phenomenological variety of neutral and ionized gas phases in the ram-pressure stripped tails of the jellyfish galaxies and also underline the complexity of this astrophysical process.

\par This paper is focused on JO204, one of the most striking jellyfish galaxies in the GASP sample. This galaxy is part of a sample of 5 GASP galaxies that have been observed with the JVLA-C in \HI. In \cite{ramatsoku2019}, we discuss another interesting jellyfish galaxy JO206 in a different environment compared to JO204. JO204 is a massive galaxy in a relatively low mass cluster and has a beautiful 30 kpc long tail of ram-pressure stripped ionized gas \citep{Gullieuszik2017}. This paper will concentrate on \HI\ observation of JO204 and the relationship between the different gas phases in relation to the star-formation activity.
\par Section 2 gives an overview of the properties and environment of JO204 as well as a brief discussion of the currently available multi-wavelength data. The JVLA \HI\ observations and data reduction are described in Section 3. Section 4 presents the analysis of the different \HI\ gas properties and the observational results of JO204. Finally, Section 5 gives a summary of the main results. 
\par In this paper, we assume a cold dark matter cosmology with $\Omega_{M}$= 0.3, $\Omega_{\Lambda}$= 0.7 and $\rm H_{0}$= 70 km$s^{-1} \rm Mpc^{-1}$. Thus, we use the same spatial scale as \cite{Gullieuszik2017} of 0.887 kpc/arcsec at z= 0.04496 for A957 \citep{biviano2017}.
% \citet{Others2013},
%and describes the problem the authors aim to solve \citep[e.g.][]{Author2012}.

\section{JO204: OVERVIEW AND PREVIOUS OBSERVATIONS}

\par
\begin{figure*}% 
  \centering
    \includegraphics[width = 180mm, height = 120mm]{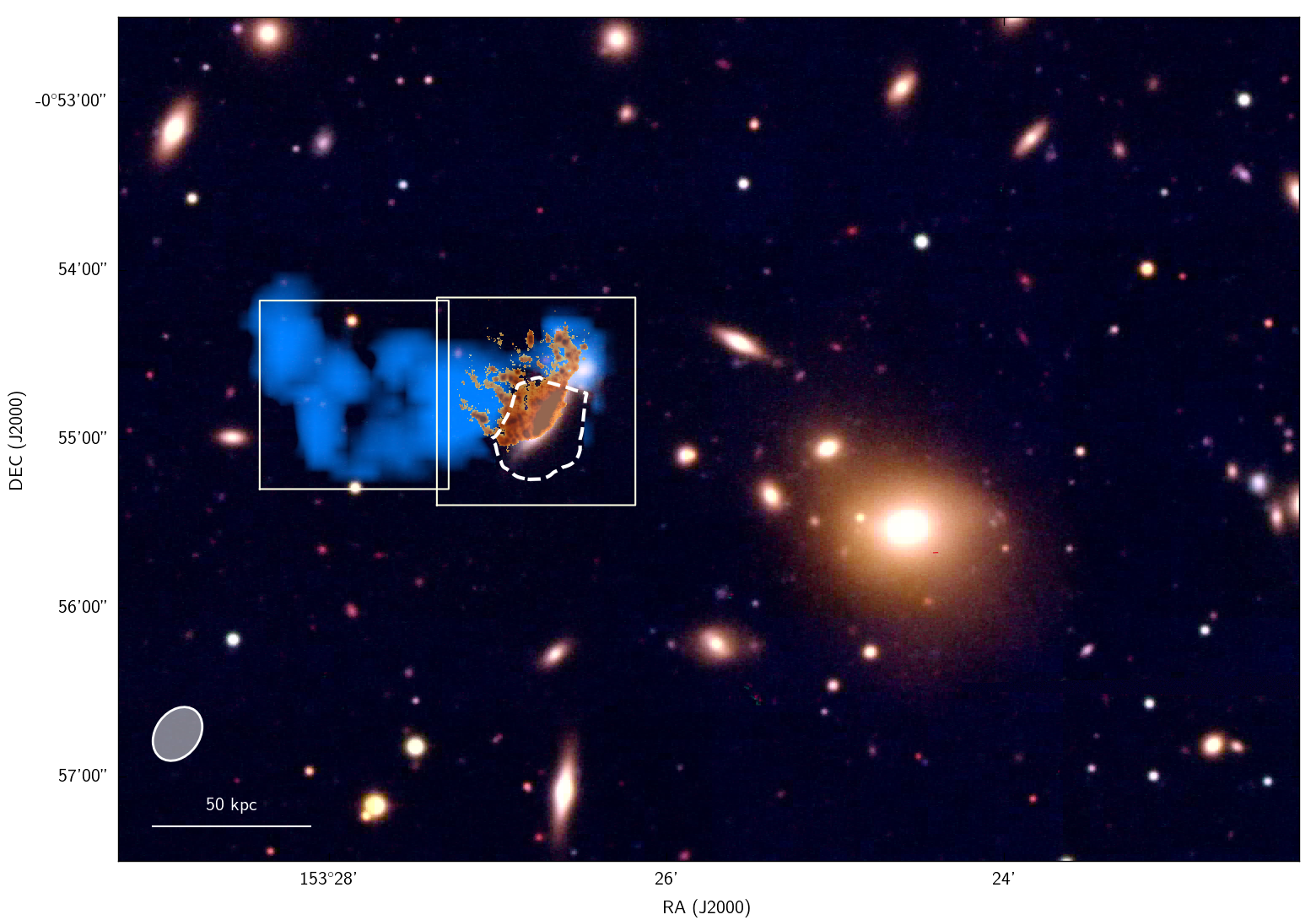}
    \caption{RGB image of central region of A957 from OmegaWINGS u- and WINGS B and V band images \protect\citep{Gullieuszik2017} overlaid with the total H$\alpha$ emission (in orange), total \HI\ emission (in blue). The dotted line is the \HI\ absorption contour (-1 mJy). The white square boxes in right and left are the old and new MUSE pointings respectively. The grey shedded ellipse in the bottom left corner is the JVLA beam size (20.7$\arcsec$ $\times$ 15.6$\arcsec$).}
    \label{JO204_RGB_HI_2}
    \end{figure*}

  JO204 ($\rm \alpha_{J2000}$= 10:13:46.84, $\rm \delta_{J2000}$=-00:54:51.27, $z=0.04243$, \citealt{Gullieuszik2017}) is a member of the Abell 957 cluster from the WINGS sample \citep{Fasano2006,Moretti2014}. It is quite a massive galaxy with a stellar mass of $\rm M_{\star}=4 \times 10^{10} \rm M_{\odot}$. The host cluster Abell 957 with a relatively low mass of $\rm M_{cl}= 4.4 \times 10^{14} \rm M_{\odot}$, has an X-ray luminosity of $\rm L_{X}= 7.8 \times 10^{43}$ ergs$^{-1}$ in the 0.1--2.4 keV band and a velocity dispersion of $\rm \sigma_{cl}$= 640 $\rm km s^{-1}$ \citep{Ebeling1996,Moretti2017}. JO204 is located 2.1 arcmin away from the cluster center (Fig \ref{JO204_RGB_HI_2}), which corresponds to a projected distance of 112 kpc. 
\par In \cite{Poggianti2016}, JO204 is classified from the WINGS B-band image as a tentative jellyfish galaxy of the highest class (JClass=5). From the projected proximity of JO204 to the cluster center and the direction of the stripped gas (opposite to the cluster center), ram-pressure by the ICM of the cluster is thought to be the most plausible mechanism. 
\par 
JO204 was observed using the MUSE IFU on the VLT on three different nights in 2015 and 2016, for a total of six 675 s exposures. Apart from a spectacular 30 kpc long tail of ionized gas, suggesting ongoing RPS (total H${\alpha}$ emission in orange, Fig \ref{JO204_RGB_HI_2}),  the MUSE data also reveal regular stellar kinematics and disturbed H${\alpha}$ gas kinematics that clearly signify a gas-only removal mechanism. Details of these observations and results are described in \citep{Gullieuszik2017}.

\par Comparing the observed ionized gas kinematics from MUSE with hydrodynamic simulations, \cite{Gullieuszik2017} found that JO204 is on its first radial infall trajectory into A957, and that ram-pressure has started stripping the ISM. The subsequent compression and collapse of the gas instigated enhanced star formation in the disk. Gas is being removed from the outer disk, bringing star-formation in the outermost regions of the disk to a halt. However, ram-pressure stripped gas has collapsed and is forming stars at a rate of 0.22  $\rm M_{\odot}/yr$ in the tail of JO204 \citep{poggianti2019}. The total SFR of JO204 is 2 $\rm M_{\odot}/yr$. Overall 13\% of the total SFR is located in the tail \citep{poggianti2019}. RPS is working outside-in and has already stripped an estimated 40\% of the total gas mass from JO204.
\par This scenario becomes more interesting given the presence of an AGN at the center of JO204. Moreover, there is also an AGN-powered extraplanar region, extending up to 15 kpc away from the stellar disk, depicting an AGN ionization cone \citep{PoggiantiNature2017,radovich2019}. 
\par Apart from optical observations, the molecular gas of JO204 was observed with the Atacama Pathfinder EXperiment (APEX) telescope. The $^{12}CO$ (2-1) transition was observed with three APEX pointings during 2016 and 2017. The three pointings covered both the main disk and the tail of JO204. A detailed description of these observations and results can be found in \cite{Moretti2018}.
\par CO line emission is detected in all three pointings, i.e. both in the main disk of the galaxy and in the H${\alpha}$ tail. Also, it is consistent with the overall distribution of the ionized gas in the central pointing, while in the two external off-disk regions, the CO line is broader than the H${\alpha}$ emission and it probably has a double component: one is coincident in velocity with the H${\alpha}$ emission, and the other is moving at a lower recession velocity.
\par There is a significant amount of CO in the central part of JO204, which corresponds to an estimated $H_{2}$ mass of 8.3 $\times 10^{9} {\rm M_{\odot}}$. The ratio of the total measured mass of molecular hydrogen and the total stellar mass is 0.42 for JO204 \citep{Moretti2018}. ALMA GASP observations of JO204 will be presented in Moretti et al. (in preparation).
\par High resolution observations with the MUSE spectrograph have yielded a very clear view of the ionized gas in the disk, tail and environment of JO204 while APEX has produced a view of the molecular gas. Understanding the neutral hydrogen content of JO204 is essential to obtain a holistic perspective on the complex interplay of the various gas phases. We already know that JO204 is falling into the cluster A957 for the first time, and experiencing RPS. Since the galaxy is on its initial infall, JO204 still contains a significant amount of ionized and molecular gas, resulting in continued star-formation activity. Hence, a considerable amount of neutral hydrogen gas is expected to be associated with JO204.

\section{ JVLA Observations and Data Reduction}

\begin{table}
\begin{tabu} to 0.5\textwidth { | X[l] |  X[l] | }
 \hline
 Property & Value  \\ [0.3ex] 
 \hline\hline
 Phase centre:$\alpha$ (J2000) & $10^{h}13^{m}46.84^{s}$  \\
  \qquad \qquad \qquad $\delta$ (J2000) & $-00^{\circ}54'51.27"$ \\
 \hline
 Central frequency & 1362.82 MHz\\
 \hline
 Bandwidth & 32 MHz\\
 \hline
 Calibrators:Primary & 3C286 \\
\qquad \qquad \quad Secondary & J1024-0052 \\
 \hline
 On source integration & 20 hrs \\
 \hline
 Observation dates & 17$^{th}$ June \& 17$^{th}$ July, 2017\\
 \hline
 RMS noise & 0.3 mJy/beam\\
 \hline
 Channel width & 6.87 $\rm km s^{-1}$ \\
 \hline
 Beam (FWHM), (P.A) & 20.7" $\times$ 15.6",-36.5$^{\circ}$ \\
 \hline
\end{tabu}
\caption{Summary of \HI\ observation}
\label{table:1}
\end{table}

\par JO204 was observed with the Jansky Very Large Array (JVLA, \citealt{Perley2009}) in its C array configuration at 1362.82 MHz, as part of the GASP project. The 15 arcsec resolution of the C array allowed us to detect  diffuse \HI\ gas in the disk and the extended tail. JO204 was observed for $\approx$ 20 hrs in total, accumulated over 5 runs on different dates. The data were collected with a total observing bandwidth of 32 MHz divided into 1024 channels. The details of those observations are summarized in Table \ref{table:1}. Due to the JVLA's large field-of-view, both JO204 and its host cluster A957 were observed.
\par The five datasets were flagged, calibrated and Fourier transformed using the NRAO Astronomical Image Processing System (AIPS, \citealt{Greisen2003} ) package. The standard calibrator 3C286 was used for flux and bandpass calibration and the nearby source J1024-0052 was used for phase calibration. Flagging was mostly conducted using the interactive SPFLG task, which displays frequency vs. time for individual baselines per polarization. Thus, radio frequency interference (RFI) for each channel in each baseline and polarization was carefully removed. Subsequently, bandpass calibration was performed, followed by amplitude calibration. Thereafter, the flagged and calibrated visibility data for the target source were split from the multi-source file and Fourier transformed. We have used Robust=1 UV weighting for imaging.
\par Both the `dirty image' and `dirty beam' (or antenna pattern) cubes for each of the five datasets were exported from AIPS. For combining the five data cubes, cleaning and further analysis, the Groningen Image Processing SYstem (GIPSY, \citealt{vdHulst1992}) software was used. Since each of the 5 corresponding channels from the five different cubes had different noise levels, different weights were used for a channel-by-channel combination of the five cubes. The channels are combined with inverse variance weights. Thus, we have a `dirty cube' with continuum emission, \HI\ line emission and \HI\ line absorption.

\par In the next step, we created masks to define the regions of continuum emission, line emission and line absorption separately. First, we  made a continuum map by visually detecting the channels without line emission or absorption and averaging them. From this continuum map, we made masks for continuum sources by visual inspection. We then subtracted the continuum map from each channel of the `dirty cube' to produce a cube with line emission and absorption only. Subsequently, we made masks for the line signals manually for each velocity channel with a flux of at least 2.5 times the rms noise at that channel, and considering coherence of the emission or absorption in adjacent channels. With the combined mask of the continuum sources and spectral line signal that we made in the previous step, we `CLEAN'-ed \citep{Hogbom1974} the `dirty' channels down to the 0.3$\sigma$ level. The clean components were restored with a 2D Gaussian beam with the same FWHM as the antennna pattern (20.7$\arcsec \times  15.6\arcsec$, $\text{P.A.}=-36.5^{\circ}$). The next task was to isolate the line signal in the cleaned cube containing both continuum sources and the line signal. For that purpose we set all pixels inside the line mask to blank and fitted a baseline to all the channels in each pixel location in the cube, thus avoiding contamination by the line signal. We then subtracted those baselines from the combined cleaned cube. Thus, we were left with a cube having only cleaned line signal that could be further analysed.

\section{\textbf{Results and} Analysis}

\subsection{\HI\ channel maps} 
\par The \HI\ channel maps show the distribution of \HI\ gas at different recession  velocities. The rest frame channel width is 6.87 $\rm km s^{-1}$, but the cube was smoothed to 20, 40 and 108 $\rm km s^{-1}$ to inspect the distribution of \HI\ gas at higher signal-to-noise ratios. The latter velocity resolution corresponds to that of the MUSE observation.
\begin{figure*}
  \centering
    \includegraphics[clip, trim=1cm 4cm 1cm 6cm, width=\textwidth]{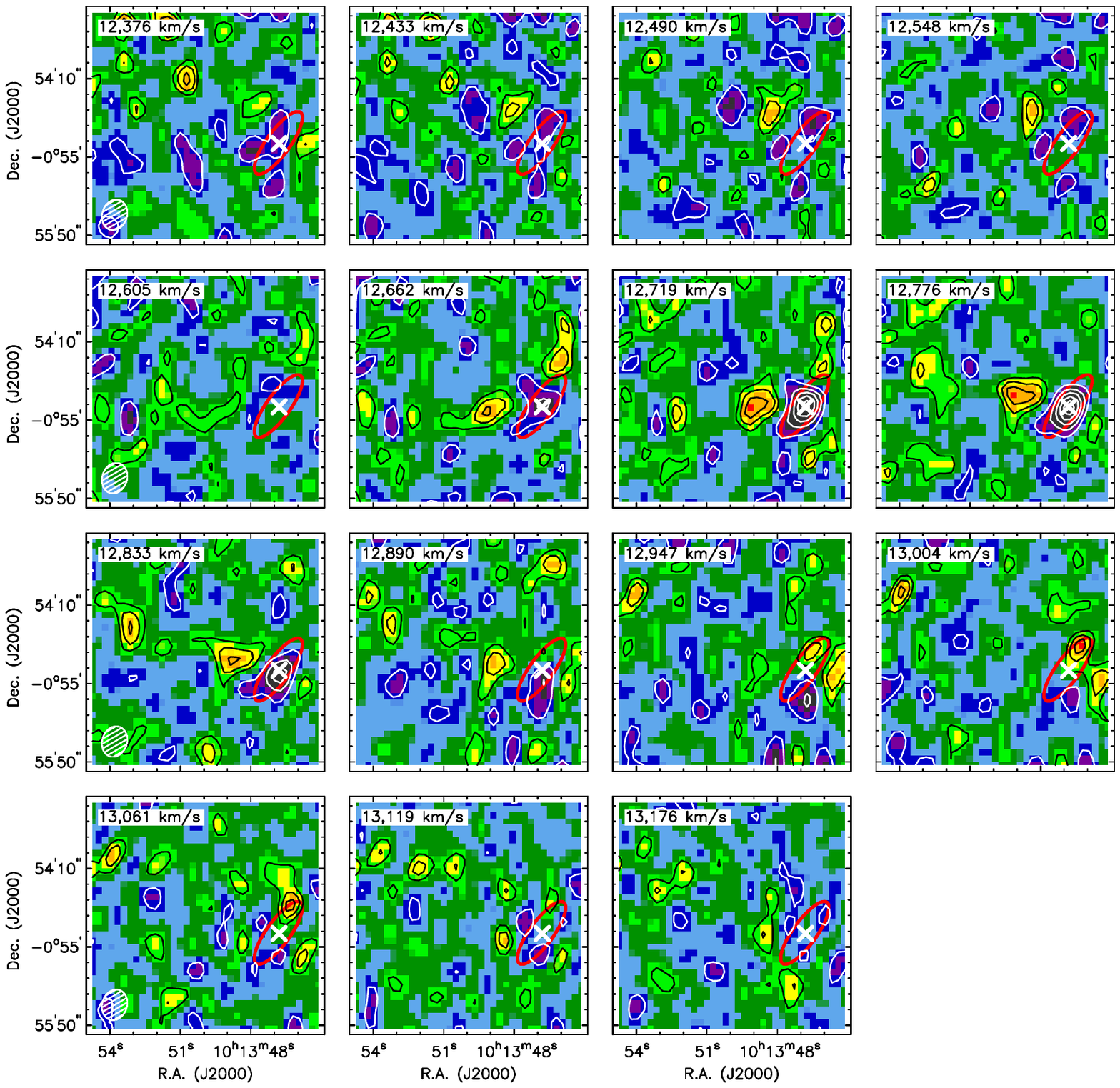}
    \caption{\HI\ channel maps smoothed to 108 $\rm km s^{-1}$: The white and black contours show the \HI\ absorption and emission, respectively. The black contours represent the \HI\ emission with contour levels of 0.25, 0.5, 1, 2 $\times$ $10^{20}$ cm$^{-2}$ and the white contours represent the \HI\ absorption with contour levels at (-1, -2, -4, -8) mJy/beam . The red ellipse is the tracer of stellar continuum. The FWHM of beam size, 20.7$\arcsec$ $\times$ 15.6$\arcsec$, is shown in the bottom left corner. The RMS noise in this cube is 0.08 mJy/beam and the colours are in 1$\sigma$ intervals.}
    \label{HI_chans_of_JO204}
    \end{figure*}
    
   Fig. \ref{HI_chans_of_JO204} shows the channel maps smoothed to 108 $\rm km s^{-1}$. The black contours represent the \HI\ emission with contour levels of (0.25, 0.5, 1, 2) $\times$ $10^{20}$ cm$^{-2}$ and the white contours represent the \HI\ absorption with contour levels at -1, -2, -4, -8 mJy/beam. The red ellipse represents the size and orientation of the stellar disk. The beam size (20.7$\arcsec$ $\times$ 15.6$\arcsec$= 18.4 kpc $\times$ 13.8 kpc) is shown in the bottom left corner. The \HI\ emission and absorption features of JO204 appear at velocities 12662<cz/km s$^{-1}$<13061. The \HI\ absorption is seen over a smaller velocity range (12662-12833 $\rm km s^{-1}$) than the \HI\ emission. The peak of \HI\ emission appears within velocities 12719-12766 $\rm km s^{-1}$ with contour levels of $\sim$5 $\sigma$.
   
   \subsection{\HI\ on H${\alpha}$ channel maps}
   \begin{figure*}% 3
  \centering
    \includegraphics[clip, trim=1cm 4cm 1cm 6cm, width=\textwidth]{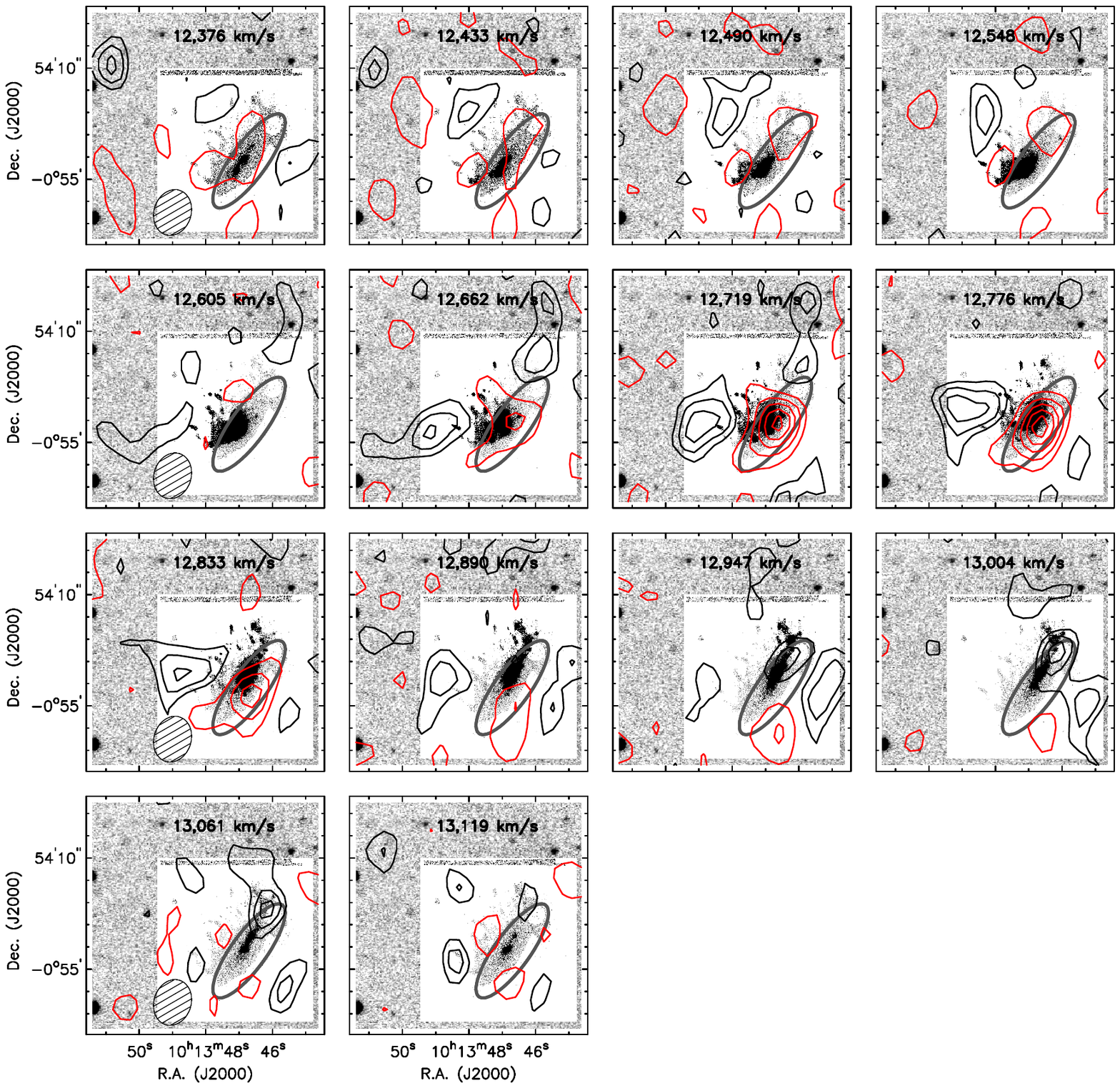}
    \caption{\HI\ on H${\alpha}$ channel maps smoothed to 108 $\rm km s^{-1}$: The red and black contours show \HI\ absorption and emission respectively. The black contours represent the \HI\ emission with contour levels of 0.25, 0.5, 1, 2 $\times$ $10^{20}$ cm$^{-2}$ and the red contours represent the \HI\ absorption with contour levels at (-1, -2, -4, -8) mJy/beam. The grey ellipse is the tracer of stellar continuum. The FWHM of beam size, 20.7$\arcsec$ $\times$  15.6$\arcsec$, is shown in the bottom left corner. The background outside the extend of the MUSE pointing is INT B-band image.}
    \label{HI_on_Ha_JO204}
    \end{figure*}
    
\par   The \HI\ emission (in black contours) and absorption (in red contours) in the channel maps is plotted on top of the H${\alpha}$ maps at corresponding velocities to compare the kinematics of these two gas phases (Fig \ref{HI_on_Ha_JO204}). At velocities between 12662 and 12833 $\rm km s^{-1}$, the \HI\ gas is offset from the H${\alpha}$ emission. Only between 12947 and 13061 $\rm km s^{-1}$, the \HI\ and H${\alpha}$ phases co-exist at the same location. At lower velocities between 12376 and 12605 $\rm km s^{-1}$, there is no \HI\ associated with the H${\alpha}$ gas. 
\par From the channel maps it is evident that the \HI\ emission extends further east relative to the H${\alpha}$ emission. The \HI\ tail is largely out of the field of view of MUSE, so the INT B-band image is shown in the regions without MUSE data. However, since, the MUSE field-of-view is much smaller than the JVLA's field-of-view, in order to investigate the presence of ionized gas in the extended part of the \HI\ tail, an additional MUSE observation has been obtained later. 
\par A region in the extended HI tail of JO204 was observed with MUSE as part of ESO Programme 0102.C-0589, a filler program designed to use idle time at the VLT UT4 during the worst weather condition. Observations were carried out on March 2-4 2019 under non-photometric conditions but with a very good seeing (between 0.4${\arcsec}$ and 1.3${\arcsec}$, as measured from the DIMM).

\par We obtained $11\times900$ sec exposures, for a total exposure time of 2.75h, which is longer that the normal one for GASP observations to compensate for possible loss due to weather conditions. These new MUSE data were reduced using the standard GASP reduction procedure (see \citealt{Poggianti2017}). We carefully inspected the resulting new datacube and could not detect any indication for the presence of ionized gas associated with the tail of JO204. We note that, since the data were taken in non-photometric nights, we used a bright star with SkyMapper photometry to confirm that the depth of the observations are similar to the one reached by the original GASP observations. The new pointing is indicated in the Fig \ref{JO204_RGB_HI_2} as the square box in the left.

    \subsection{Total \HI\ map}
  
  \begin{figure}% 3
  \centering
    \includegraphics[clip, trim=2.4cm 9cm 4cm 4.9cm, width=0.45\textwidth]{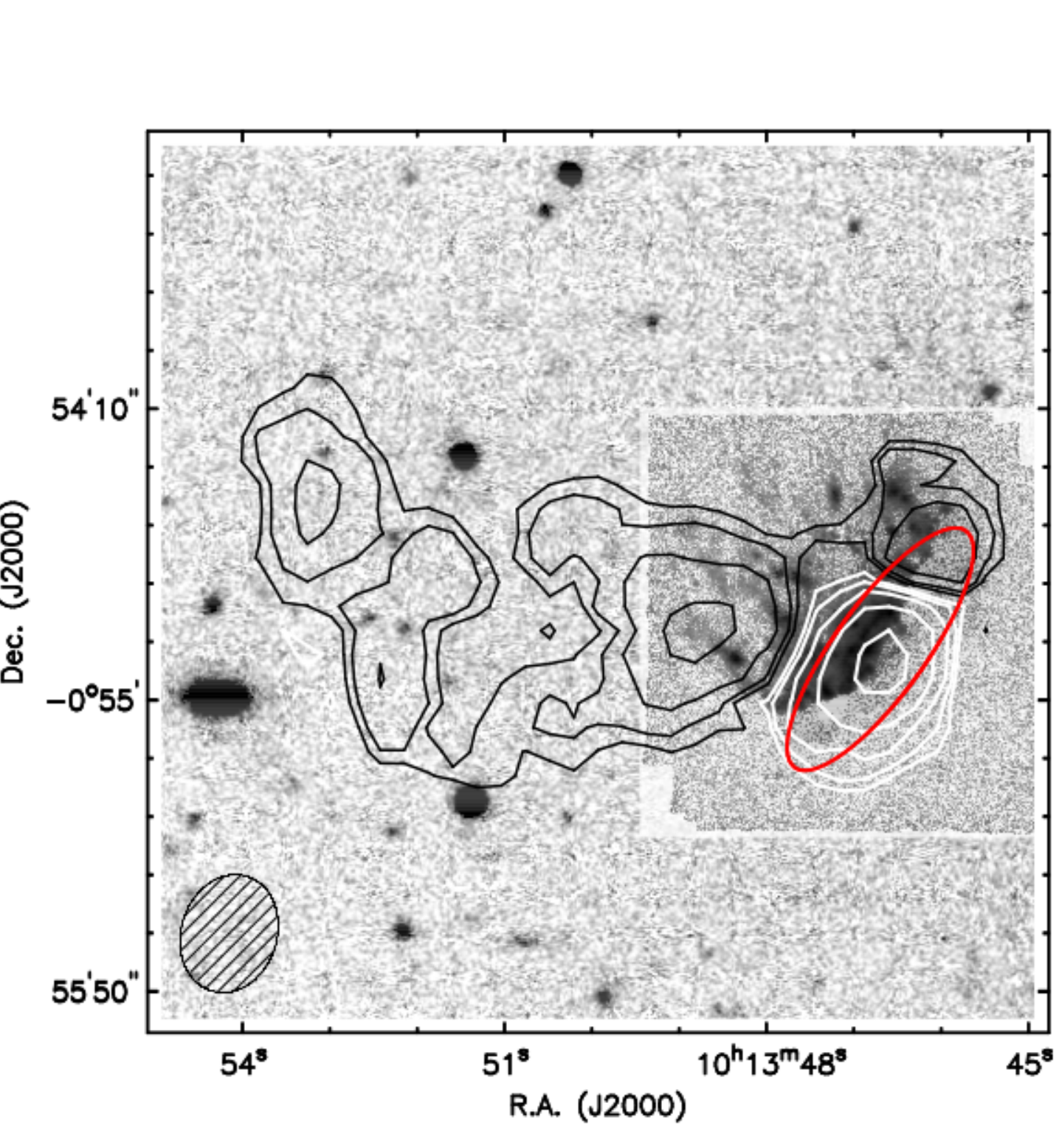}
    \caption{The JVLA \HI\ column density contours are overlaid on MUSE H$\alpha$ map: black contours are \HI\ emission with contour levels at (0.25, 0.5, 1, 2) $\times$ $10^{20}$ cm$^{-2}$; white contours represent \HI\ absorption with contour levels at (-1, -2, -4, -8) mJy/beam. Outside the extent of the MUSE pointing, the background is the INT B-band image. The red ellipse is the tracer of the stellar disk. The FWHM beam size of 20.7$\arcsec$ $\times$ 15.6$\arcsec$ is in the bottom left of the plot. }
    \label{HI_map_of_JO204}
    \end{figure}
    
\par The total \HI\ map shows the integrated column density of the \HI\ gas. Since there is both emission and absorption in the \HI\ cube, 3-dimensional masks were made that encompass both \HI\ emission and absorption regions in the cube. The masks are made manually for each velocity channel seen in \HI\ emission and absorption with a \HI\ flux of at least 2.5 times the rms noise at that channel, and visually determining some consistency in the distribution of the gas compared to the neighbouring channels. Pixels outside this mask were set to zero and pixels inside the mask were summed up along the frequency axis to obtain the total \HI\ map. By this method, an optimal signal-to-noise ratio can be obtained for each pixel in the \HI\ map. It is impossible to define the 3$\sigma$ column density level as the noise varies across the map because a different number of channels was added for different pixels in the map \citep{verheijen2001}. The conversion from pixel values (Jy/beam) to \HI\ map column density (i.e., cm$^{-2}$) was done using:
\begin{equation}
\rm N_{HI}=1.83 \times 10^{18} (1+z) \int T_b d\nu
\end{equation}
where $\rm N_{HI}$ is the \HI\ column density in cm$^{-2}$, $\rm d\nu$ is the velocity width in $\rm km s^{-1}$ over which the emission line is integrated at a pixel in the map and $\rm T_{b}$ is the brightness temperature in Kelvin that is calculated using:

\begin{equation}
 \rm T_{b}=\frac{605.7}{\Theta_{x} \Theta_{y}} S_{\nu} {\Bigg(\frac{\nu_{0}}{\nu}\Bigg)}^{2}
\end{equation}

\noindent where $\rm S_\nu$ is the flux density in mJy/beam, $\rm \nu_{0}$ and $\rm \nu$ are the rest frequency and the observed frequency of the \HI\ emission line respectively, and $\rm \Theta_{x}$ and $\rm \Theta_{y}$ are the major and minor axes of the Gaussian beam in arcseconds. For JO204, the average $\rm N_{HI}$ sensitivity at the $3\sigma$ level is $5 \times 10^{19}$ cm$^{-2}$.
\newline The total mass of \HI\ gas (in $\rm M_\odot$ ) seen in emission is determined using:
\begin{equation}
\rm M_{HI}=\frac{2.36 \times 10^5 D_{L}^2}{(1+z)} \int S_\nu d\nu
\end{equation}
where $\rm \int S_\nu d\nu$ is the integrated \HI\ flux in Jy $\rm km s^{-1}$ and $\rm D_{L}$ is the luminosity distance to the galaxy in Mpc.
\newline The \HI\ absorption column density is calculated using:
\begin{equation}
\rm N_{HI}^{abs}=1.83 \times 10^{18} \times T_{spin} \int \lambda d\nu
\end{equation}
where $\rm N_{HI}^{abs}$ is the \HI\ absorption column density, $\rm T_{spin}$ is the assumed spin temperature of 100 K and
\begin{equation}
\rm \int \lambda d\nu= \sum_{i=1}^{n} -ln\Bigg(1-\frac{f_i}{S}\Bigg) \times \Delta \nu
\end{equation}
where $\rm f_i$ is the flux density (in Jy) of the \HI\ absorption at frequency channel i, S is the continuum flux (in Jy) outside the absorption profile. $\rm \Delta \nu$ is the width of each channel in $\rm km s^{-1}$ and n is the number of channels where the \HI\ absorption is seen. 

\par In the \HI\ map of JO204 we also see strong \HI\ absorption because some of the \HI\ gas is located in front of the radio continuum source (the central AGN of JO204 in this case). 

\par Fig \ref{HI_map_of_JO204} shows the \HI\ contours overlaid on the H$\alpha$ map from MUSE, which is inset in a B-band image from the INT. The black contours represent the \HI\ emission with contour levels of (0.25, 0.5, 1, 2) $\times$ $10^{20}$ cm$^{-2}$ and the white contours represent the \HI\ absorption with contour levels at -1, -2, -4, -8 mJy/beam . The FWHM of the beam is 20.7$\arcsec$ $\times$ 15.6$\arcsec$, shown in the bottom left corner. The \HI\ tail extends eastward up to 90 kpc from the stellar body of the galaxy, much further than the H$\alpha$ tail which extends only 30 kpc.  The \HI\ absorption in the central region of the galaxy makes it difficult to estimate the total amount of \HI\ gas in the disk of JO204.
The total amount of \HI\ gas seen in emission is  (1.32 $ \pm 0.13) \times 10^{9} \rm M_{\odot}$. However, this is not the total \HI\ mass of JO204 due to the \HI\ absorption, which corresponds to a  column density of 3.2 $\times 10^{20}$ cm$^{-2}$, assuming a spin temperature of $\rm T_{spin}$= 100 K.
\par Due to the strong \HI\ absorption signal against the central continuum source and the relatively large synthesized beam compared to the disk of the galaxy, we can not measure the amount of neutral hydrogen gas associated with the disk of the galaxy. Therefore,  in the case of JO204, we are unable to compare the star formation activity (traced by the ionized gas) in the disk and in the tail with the amount of cold neutral hydrogen gas.

 \subsection{\HI\ global profile}

  \begin{table*}
\begin{tabu} to 1.05\textwidth {| X[l] | X[l] | X[l] | X[l] | X[l] |} 
 \hline
 Survey & Unit & GASP & NVSS & FIRST  \\[0.3ex] 
 \hline\hline
 RA & J2000 & $10^{h}13^{m}46.84^{s}$ & $10^{h}13^{m}46.93^{s} \pm 0.11^{s}$ & $10^{h}13^{m}46.845^{s}$ \\ 
 \hline
 DEC & J2000 & $-0^{\circ}54'50.7\arcsec$ & $-0^{\circ}54'49.2\arcsec \pm 1.8\arcsec$ & $-0^{\circ}54'51.1\arcsec$\\
 \hline
 Measured major \& minor axis, P.A. & arcsec, arcsec \& deg & 21.51 $\pm 0.4$, 17.11 $\pm 0.52$, 165 & & 6.8, 5.9, 159.9 \\
 \hline
 Inferred intrinsic  
 size of major \& 
 minor axis, P.A. & arcsec, arcsec \& deg & 5.78, 6.94 & <38.3, <34.3 & 3.04, 1.2, 121.1 \\
 \hline
 Integrated flux & mJy & 11 $\pm$ 0.2 & 11.3 $\pm$ 0.6 & 7.97 $\pm$ 0.15 \\[0.3ex] 
 \hline
\end{tabu}
\caption{Summary of different continuum observations of JO204. For the FIRST survey, individual sources have 90\% confidence error circles of <0.5$\arcsec$ at the 3 mJy level.}
\label{table:2}
\end{table*}

\begin{figure*}% 3
  \centering
    \includegraphics[width=\textwidth]{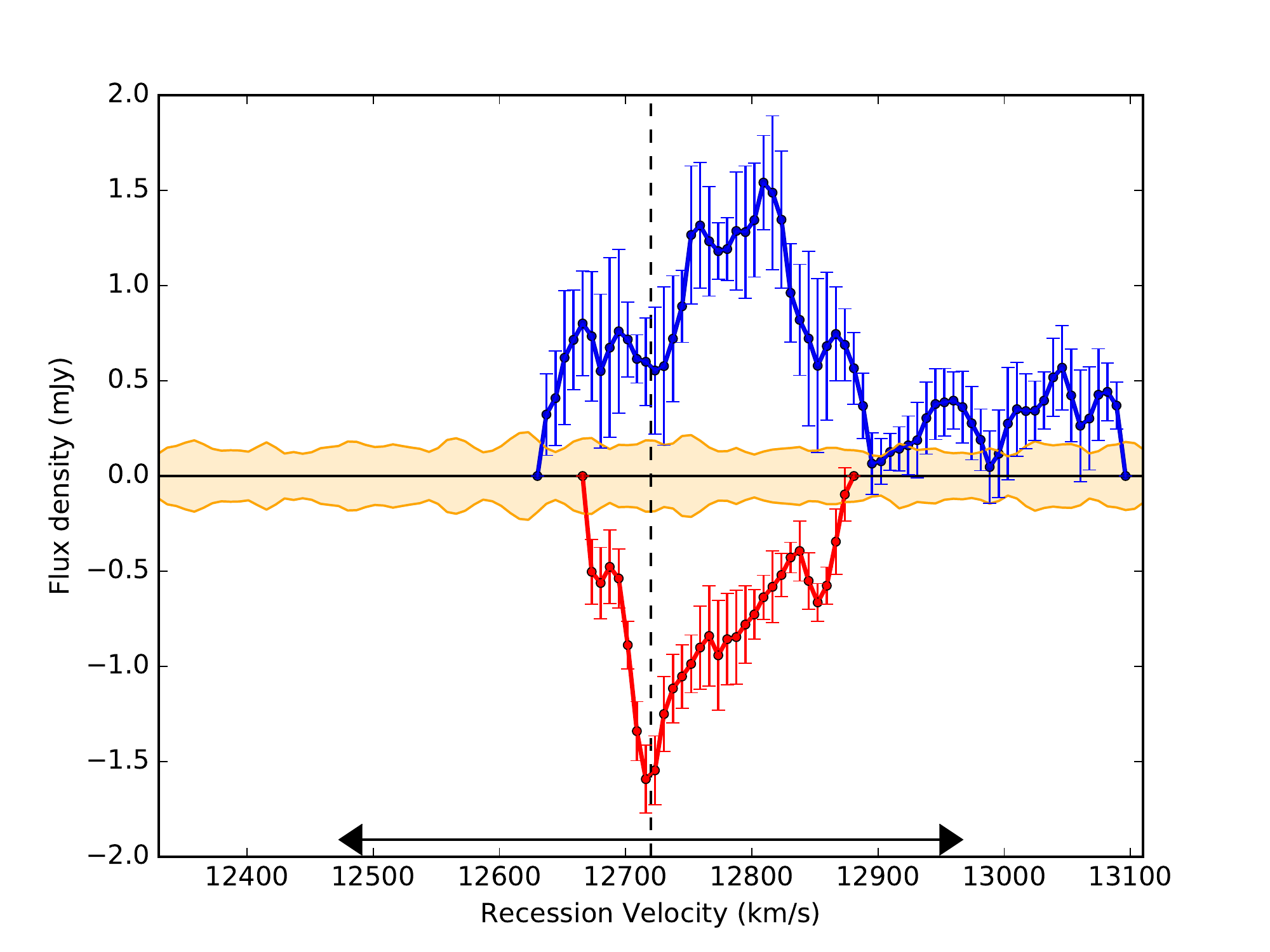}
    \caption{\HI\ global profiles: blue and red profiles are emission and absorption profiles, respectively. The dotted in the middle is the systemic velocity (12720 $\rm km s^{-1}$) based on the stellar component at the center of the galaxy. The horizontal line with arrows on either side is a representative of the width of global profile as expected from the Tully-Fisher (TF) relation. We have added a yellow band, showing the rms noise as a function of recession velocity. We have considered region of single beam size to calculate the rms as a function of recessional velocity.}
    \label{global_profile}
    \end{figure*}

\begin{figure}% 3
  \centering
    \includegraphics[clip, trim=0cm 5cm 0cm 5cm,width=0.5\textwidth]{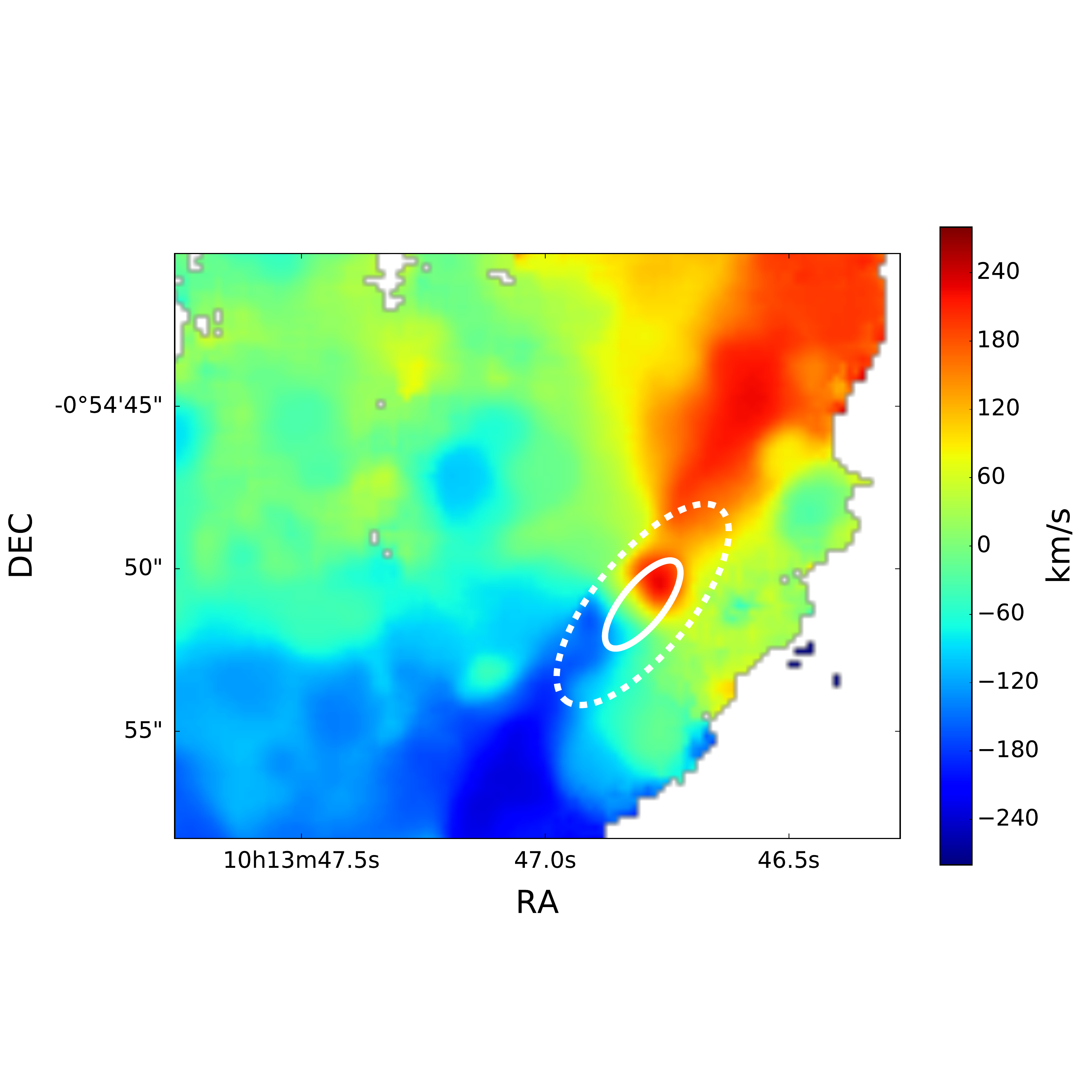}
    \caption{The radio continuum sources overlaid on H$\alpha$ velocity field. The solid and dotted white ellipses at the center represents the estimated size of the continuum source based on the FIRST and GASP observations respectively. The colour bar is velocity of H$\alpha$ gas in $\rm km s^{-1}$. Evidently, there is a steep velocity gradient across the radio continuum source.}
    \label{vfield_cont_src}
    \end{figure}

 \par   Fig \ref{global_profile} illustrates the global profile of \HI\ emission and absorption for JO204, which shows the \HI\ flux density as a function of velocity for emission (top panel) and absorption (bottom panel). The vertical dashed line indicates the systemic velocity of cz=12,720 $\rm km s^{-1}$ based on z=0.04243 as reported in \cite{Gullieuszik2017}, and derived from the stellar component at the center of the galaxy. The horizontal line with arrows on either side illustrates the width of the global profile as expected from the Tully-Fisher (TF) relation \citep{tully-fisher1977}. The total K$_{s}$-band magnitude (m$_{K_s}$=12.08 $\pm$ 0.081) of JO204 from the Two-Micron All Sky Survey (2MASS, \citealt{2mass2006}) and a luminosity distance of D$_L$=188 Mpc imply a total absolute K$_{s}$-band magnitude of -24.3 mag. Inserting into the K$_{s}$-band TF relation by \cite{ponomareva2017} yields an expected intrinsic line width of W$_{50}$=480 $\rm km s^{-1}$, consistent with the K$_{s}$-band TF relation by \cite{verheijen2001}. We estimate the inclination of the galaxy at i=75$^\circ$ based on the observed ellipticity of the outer isophote and an assumed intrinsic thickness of the stellar disk of q$_{0}$=0.15, which results in an expected observed line width of W$_{50} \times$ Sin(i)=465 $\rm km s^{-1}$.

    \par The emission profile is asymmetric with respect to the systemic velocity. The redshifted \HI\ emission corresponds to the northern, receding part of \HI\ gas disk that is not yet fully stripped as can be seen in Fig \ref{HI_map_of_JO204}. Given the systemic recession velocity of cz=12,720 $\rm km s^{-1}$ of JO204 and the projected maximum rotational velocity of 464/2=232 $\rm km s^{-1}$, we would not expect to see HI emission at an observed recession velocity in excess of cz=12720+232=12,952 $\rm km s^{-1}$. Thus, some \HI\ emission in the receding part is observed at higher recession velocities than expected from regular rotation (12965-13100 $\rm km s^{-1}$). As mentioned before, this redshifted \HI\ gas corresponds to the northern part of the galaxy and is already displaced from the rotating stellar disk due to ram-pressure. The corresponding and missing blue-shifted emission would have come from the southern part of the galaxy that is already in an advanced stage of RPS. As a result of the removal of \HI\ gas from the southern part of JO204, the extension of \HI\ emission toward lower velocities is less than expected from the TF relation. 
    
    \par The absorption profile is also asymmetric with an extended redshifted wing. We also see blueshifted \HI\ absorption, but that is not so extended in velocity. The \HI\ absorption is caused by atomic hydrogen between the observer and a background radio continuum source, in this case located in the center of JO204. Redshifted \HI\ absorption is caused by gas that moves towards the continuum source, which can be explained in two ways. In the first scenario, ram-pressure is pushing the gas towards the AGN, creating this redshifted absorption and triggering the nuclear activity. Hence the redshifted \HI\ absorption wing lends support to the findings by \citealt{PoggiantiNature2017} who found a strong correlation between RPS and the presence of AGN activity. In the second scenario, the redshifted wing of the absorption profile could be due to the very fast rotation of the \HI\ gas disk seen partially in front of an extended continuum source. In Fig \ref{vfield_cont_src} (solid white ellipse), we see that ionized gas is moving over a large range of velocities in front of the small radio continuum source as derived from FIRST observations (FIRST survey, \citealt{FIRST1994}, \citealt{FIRST1995}) as well as in front of the more extended star-forming region of the radio continuum source that is derived from our JVLA-C \HI\ observation (dotted white ellipse in Fig \ref{vfield_cont_src}). In this latter scenario, the redshifted \HI\ absorption wing is not necessarily evidence of gas being pushed towards the AGN. If the width of the modelled absorption profile is wider than the observed profile, then the absorption can arise in the rotating \HI\ disk. However, if the red wing of the observed absorption profile falls outside the modelled profile, this would be a strong hint for gas being pushed towards the AGN. In order to investigate the plausibility of these scenarios, we make an attempt to model the \HI\ absorption line profile in the following subsection. 
    
    \subsection{Modelling the \HI\ absorption profile}

    \par Ionized gas (H${\alpha}$ or H${\beta}$) emission from the MUSE data shows two components, a broad component due to the central AGN outflow and a narrow component associated with a rotating gas disk. Assuming similar kinematics of the ionized and \HI\ gas and a uniform \HI\ gas column-density in the central region of JO204, we have modelled the \HI\ absorption profile based on the MUSE emission line data. From studies of more nearby galaxies with ionized and neutral Hydrogen gas (e.g. \citealt{martinsson2016}), it is evident that the neutral and ionized phases of the Hydrogen gas follow the same kinematics. This is true across the Hubble sequence. We have seen this to be the case also in E/S0 galaxies in \cite{morganti2006} and, with a larger sample, \cite{serra2014} -- and there are more individual cases in the literature. Usually, \HI\ and H$\alpha$ are correlated unless there is some major feedback episode that is only accelerating warm and hot gas which is not the case for JO204 as we understand from the BPT diagram \citep{Gullieuszik2017}. The MUSE data has much better spatial resolution ($\approx$1 arcsec) compared to the JVLA-C resolution ($\approx$15 arcsec) although the velocity resolution is worse ($\approx$108 $\rm km s^{-1}$) in comparison with the JVLA-C ($\approx$7 $\rm km s^{-1}$). For our \HI\ absorption modelling, we have created a model datacube with a spatial sampling that is the same as the MUSE cube (0.2${\arcsec}$) and a velocity sampling that is the same as the JVLA C datacube (7 $\rm km s^{-1}$).

 \par We have information about the radio continuum source from 3 different surveys with different resolutions as summarized in Table \ref{table:2}. The FIRST survey with a 5$\arcsec$ beam has a typical rms noise of 0.15 mJy/beam or 3.8 K while the FIRST catalog has a detection threshold of 0.98 mJy/beam. The FIRST peak flux is 6.88 mJy/beam and the total integrated flux is 7.97 mJy/beam, indicating that the source might be slightly extended.
The NVSS survey (\citealt{nvss1998}) with a 45$\arcsec$ beam has a typical rms noise of 0.45 mJy/beam or 0.14 K. Consequently, the FIRST survey is three times more sensitive to point sources compared to the NVSS, but the NVSS survey is 27 times more sensitive for extended emission than the FIRST survey. We infer that the radio continuum flux is a combination of contributions from the central AGN in JO204 and an extended star forming region around it. Since the FIRST observation could only detect 8 mJy from the brightest central region of the source, we infer that most of the flux (8 mJy out of 11 mJy) is coming from the central AGN region of the source, while an additional (11-8=)3 mJy is coming from a more extended star forming region that JVLA-C and NVSS observations could detect. Thus, the difference of 3 mJy between the FIRST and NVSS or GASP can be due to the fact that the brightness temperature sensitivity of the FIRST survey is significantly lower for extended emission than that of the NVSS or GASP survey. Of course, a 3 mJy difference in flux measurement between these surveys may also be caused by the variability of the AGN i.e. the radio continuum source between the time of these three different
observations. The position of the continuum source from FIRST is well-defined at $\alpha$=10:13:46.85, $\delta$=-00:54 51.1 while the intrinsic size of 3.0$\arcsec$ $\times$ 1.2$\arcsec$ and position angle of 121$^\circ$ are estimated from a deconvolution of the observed source with the synthesized beam.

 \begin{figure}% 3
  \centering
    \includegraphics[clip, trim=2.5cm 4.6cm 1cm 6.3cm, width=0.45\textwidth]{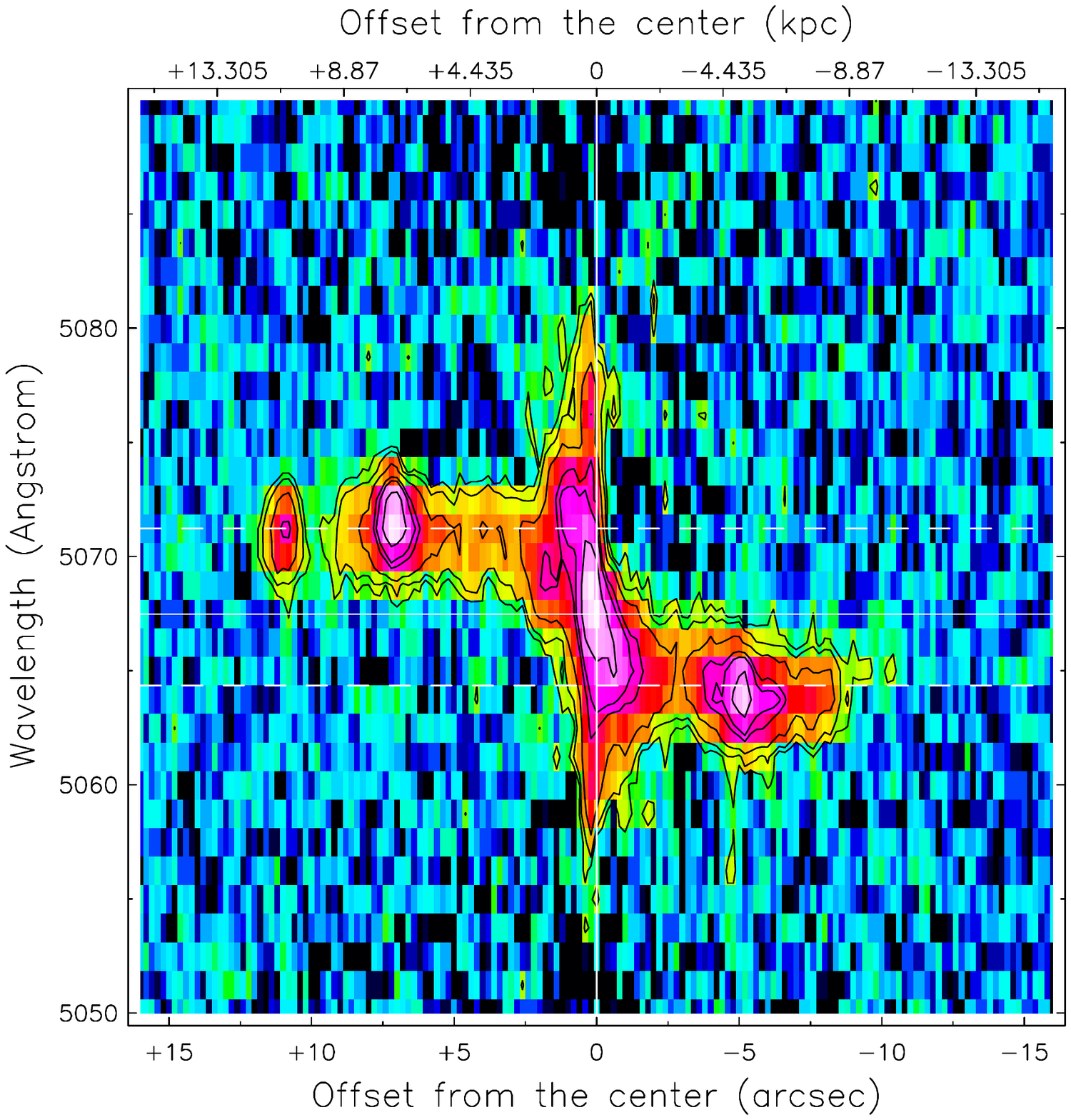}
    \caption{Position velocity (PV) diagram of the H${\beta}$ line along the major axis of JO204 (position angle= -35$^{\circ}$). The vertical solid line is through the center and the horizontal solid line indicates the systemic velocity of JO204. There are two components of this line: one broad-line region associated with the AGN outflow and another narrow-line region that is coming from the regular rotating gas disk. The dashed horizontal lines represents the extend in velocities from TF relation.}
    \label{HB_PV}
    \end{figure}

 \begin{figure*}% 3
  \centering
    \includegraphics[clip, trim=2cm 7.5cm 2cm 5cm, width=0.9\textwidth]{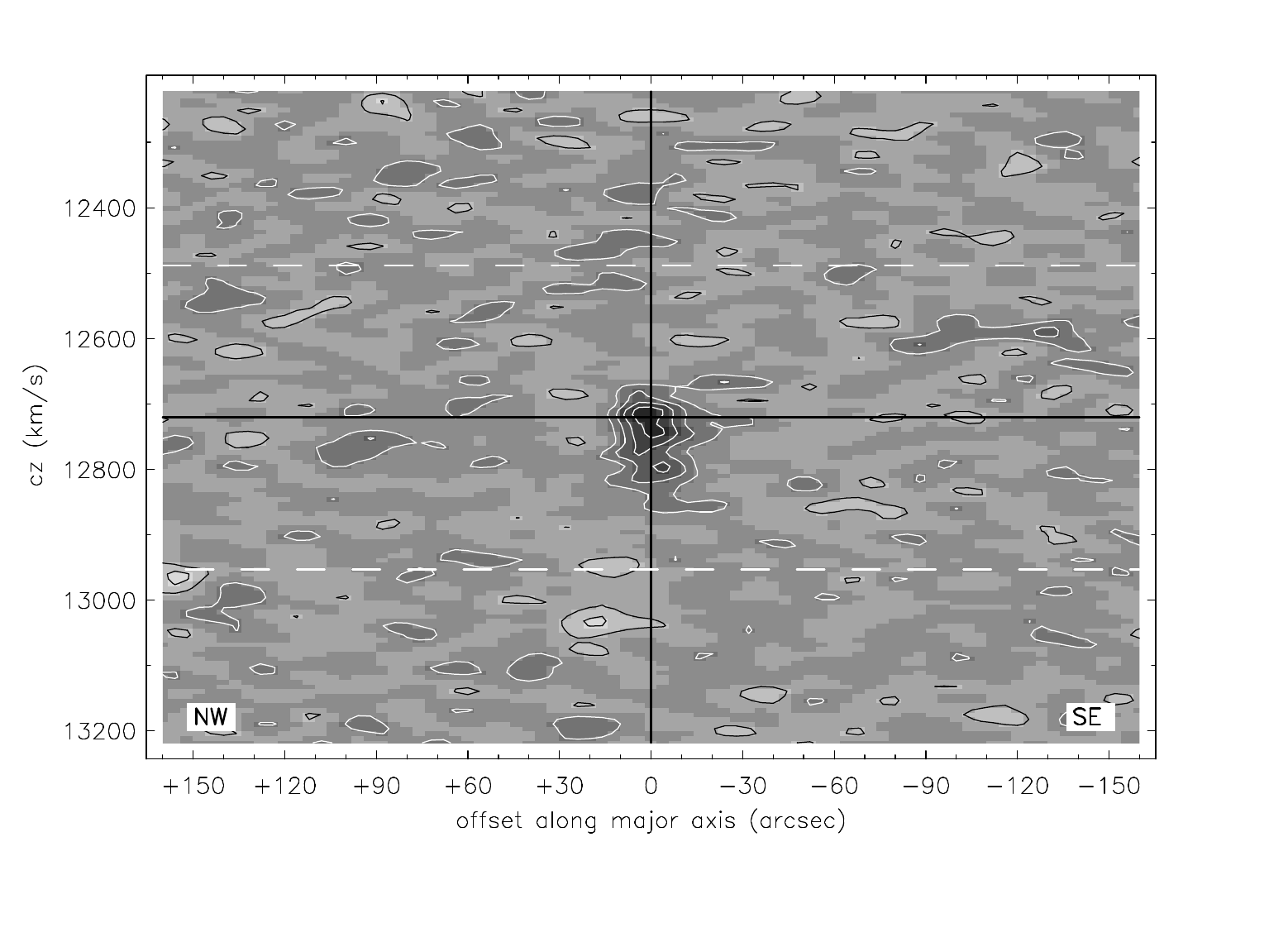}
    \caption{Position velocity (PV) diagram of the \HI\ line along the major axis of JO204 (position angle= -35$^{\circ}$). The vertical solid line is through the center and the horizontal solid line indicates the systemic velocity. Contours are in steps of 1.5$\sigma$; white is absorption, black is emission. The noise contours indicate the size of the beam. The dashed white horizontal lines represents the extend in velocities from TF relation.   }
    \label{HI_PV}
    \end{figure*}

    \begin{figure*}% 3
  \centering
    \includegraphics[width=\textwidth]{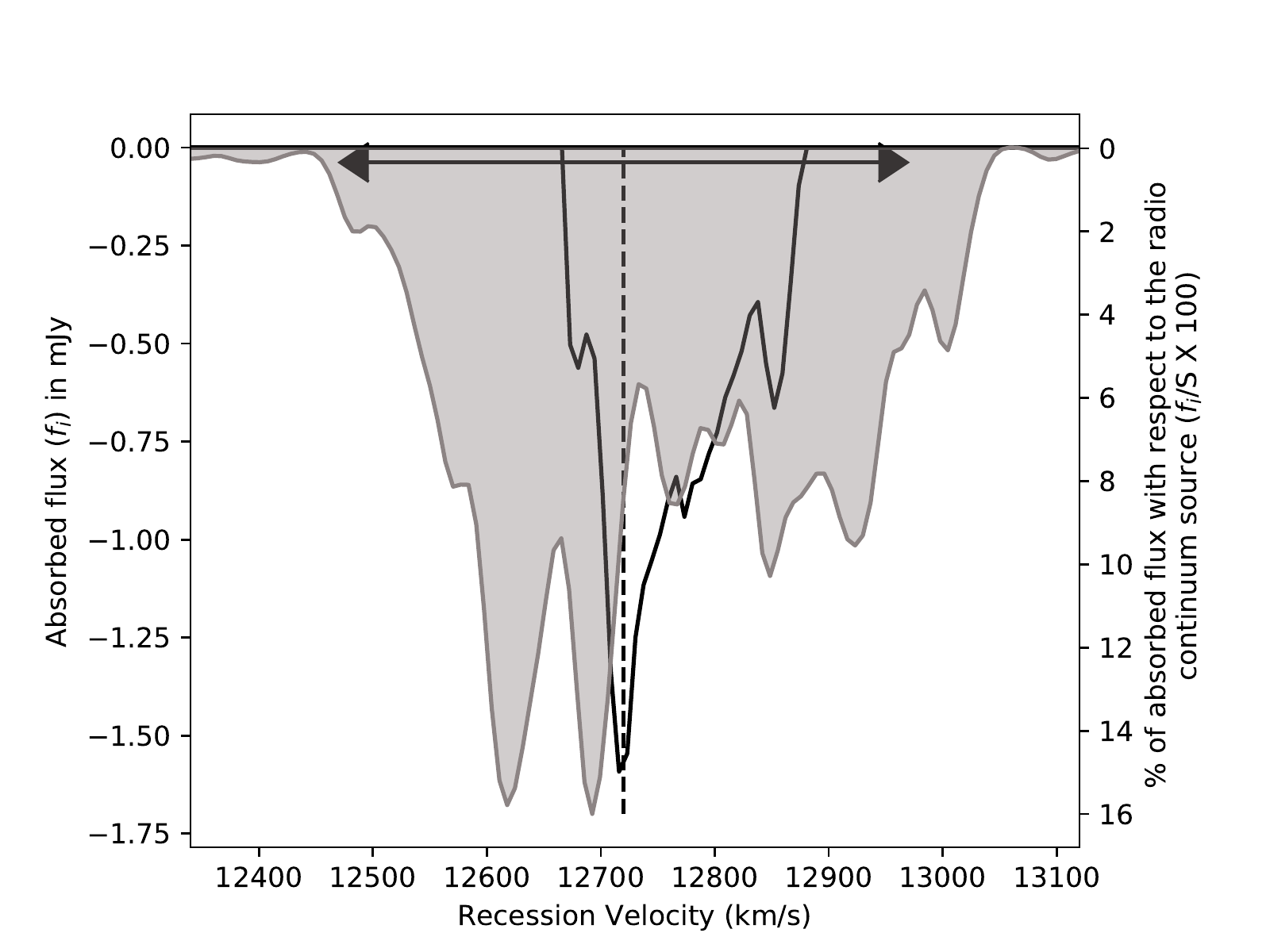}
    \caption{Absorption profiles: modelled \HI\ absorption (grey area, produced using H${\beta}$ line) and observed \HI\ (black line) absorption. The dotted line in the middle is the systemic velocity (12720 $\rm km s^{-1}$) deduced from the MUSE observation. The horizontal line with arrows on either side depicts the width of global profile from the TF relation. On the left side of the y-axis we show absorbed flux or $\rm f_{i}$ in mJy. On the right side of the y-axis we show the percentage of absorbed flux with respect to the radio continuum source ($\rm f_{i}$/S $\times$ 100.) }
    \label{abs_prof_HB_model_HI}
    \end{figure*}

 \par Adopting the estimated size of the continuum source in the inner region based on the FIRST observation, we investigate the \HI\ gas kinematics in front of that small region. If we make the plausible assumption that the \HI\ and H${\alpha}$ gases have the same kinematics in the central region, we can consider the line-of-sight velocities of the H$\alpha$ gas as provided by the MUSE data across the small 3.0$\arcsec$ $\times$ 1.2$\arcsec$ region of the continuum source . The H${\alpha}$ line, however, is very close to the [NII]6548 and [NII]6583 lines. There are two components for each of these lines: a broad component in the very central region associated with the AGN and a narrow component that is emanating from the regular rotating gas disk. The superposition of the broad component of these three lines causes difficulty to disentangle the H${\alpha}$ narrow and broad components from the NII lines.
 
 \par To avoid using the H${\alpha}$ line, which is contaminated by the broad components of the NII lines, we take advantage of the H$\beta$ emission line. H${\alpha}$ and H${\beta}$ emission represent the kinematics of the same gas. Fig \ref{HB_PV} shows the position-velocity diagram (PV diagram), a cut through the galaxy center along the major axis of the optical disk (P.A.=-35$^{\circ}$), which provides a fair indication of the kinematics along the major axis. There are two components for the H$\beta$ line as well: a broad component in the very central region associated with the AGN that is seen along the drawn vertical line, and a narrow component that is emanating from the regular rotating gas disk which is also extended horizontally on either side of the center. So, instead of the H${\alpha}$ emission, we have used the H${\beta}$ line to model the \HI\ absorption profile. In the Fig \ref{HB_PV}, the dotted lines represent the velocity width as expected from the TF relation.  Clearly, the ionized gas has reached the maximum rotation velocity as expected from TF relation already in the central region.
 
 Fig \ref{HI_PV} is an \HI\ PV-diagram through the center of JO204 at a position angle of -35$^{\circ}$ and at a velocity resolution of 20 $\rm km s^{-1}$. The white contours represents absorption while the black ones represents emission. The contours are drawn in steps of 1.5$\sigma$.  The horizontal black line indicates the systemic velocity cz=12720 $\rm km s^{-1}$. The vertical black line indicates the center of the galaxy. The horizontal, dashed white lines indicate the expected TF line width. So, the \HI\ in absorption covers a smaller velocity range than expected from TF relation, i.e. regular rotation while from Fig \ref{HB_PV} we see that the ionized gas nicely fits the expected TF width. The 3$\sigma$ peak of \HI\ emission is roughly at +20$\arcsec$ and 13020 $\rm km s^{-1}$ which is in the outer disk at the NW side of the galaxy. Moreover, it recedes at a velocity that is larger than the TF expected rotation velocity.
 
 \par Since our main objective is to model the \HI\ gas kinematics based on the narrow component of the ionized gas, which we assume to represent the rotating gas disk, a double Gaussian (for both the narrow and broad components) is fitted to the observed H${\beta}$ line profiles in the MUSE cube. A \HI\ model-cube is constructed using the velocity centroids of the fitted narrow Gaussian component. In addition, the amplitude is set at a constant value since a uniform column-density is assumed for the \HI\ phase. Ionized gas usually shows a much larger velocity dispersion than the cold neutral gas. Therefore, instead of adopting the velocity dispersion of the ionized gas ($\sim$ 45-70 $\rm km s^{-1}$), which would result in a much broader velocity profile, we have assumed a constant velocity dispersion of 10 $\rm km s^{-1}$ for our model \HI\ cube (\citealt{petric2007}, \citealt{boosma}).
 
 \par The resultant model \HI\ absorption profile depends on the angular distribution of the radio continuum source and the column density of the \HI\ gas in front of the extended radio continuum source. So, after multiplying the model \HI\ cube with a 2D Gaussian model for the radio continuum source, with a size and flux given by the FIRST observation, and subtracting the result from the model radio continuum source, we have obtained the model \HI\ absorption profile as depicted by the grey area in Fig \ref{abs_prof_HB_model_HI}). In this figure, the observed \HI\ absorption profile is drawn with a solid black curve. Since we are interested in the width and shape of the modelled absorption profile, we have scaled the depth of the modelled profile to the depth of the observed absorption profile. It should also be noted that the structure in the modelled absorption profile is a reflection of kinematic irregularities of the observed ionized gas disk since we have assumed a uniform column density.
 
 \par Though both of these profiles are asymmetric, the width of the model \HI\ absorption profile is larger than the observed \HI\ absorption profile. This could be a consequence of the likely clumpy nature of the \HI\ gas disk in the very center of JO204. In other words, the narrower observed absorption profile is consistent with a clumpy distribution of the \HI\ gas such that not all recession velocities of the H$\beta$ velocity field seen in front of the continuum source are sampled by sufficiently high column density \HI\ gas to give rise to absorption at those velocities.
 Indeed, the assumption of a uniform column-density of the \HI\ gas might be inappropriate. In reality, the observed \HI\ absorption profile may have come from a collection of discrete \HI\ gas clumps in front of the small radio continuum source only at certain velocities, and thus might not sample the entire velocity range displayed by the H${\beta}$ velocity field. Furthermore, we have to consider that in our model, all of the H${\beta}$ gas that is assumed to be representative of the kinematics of the \HI\ gas is actually in front of the radio continuum source. In reality, not all of the observed ionized gas may be on circular orbits and uniformly fill the midplane of the galaxy. This may contribute to the asymmetry in the modelled \HI\ absorption profile. Finally, we note that the width of the modelled absorption profile is similar to the rotational velocity expected from the TF relation, indicating that JO204 reaches its maximum rotational velocity in the very center of the galaxy.

\begin{figure*}
  \centering
    \includegraphics[width = \textwidth]{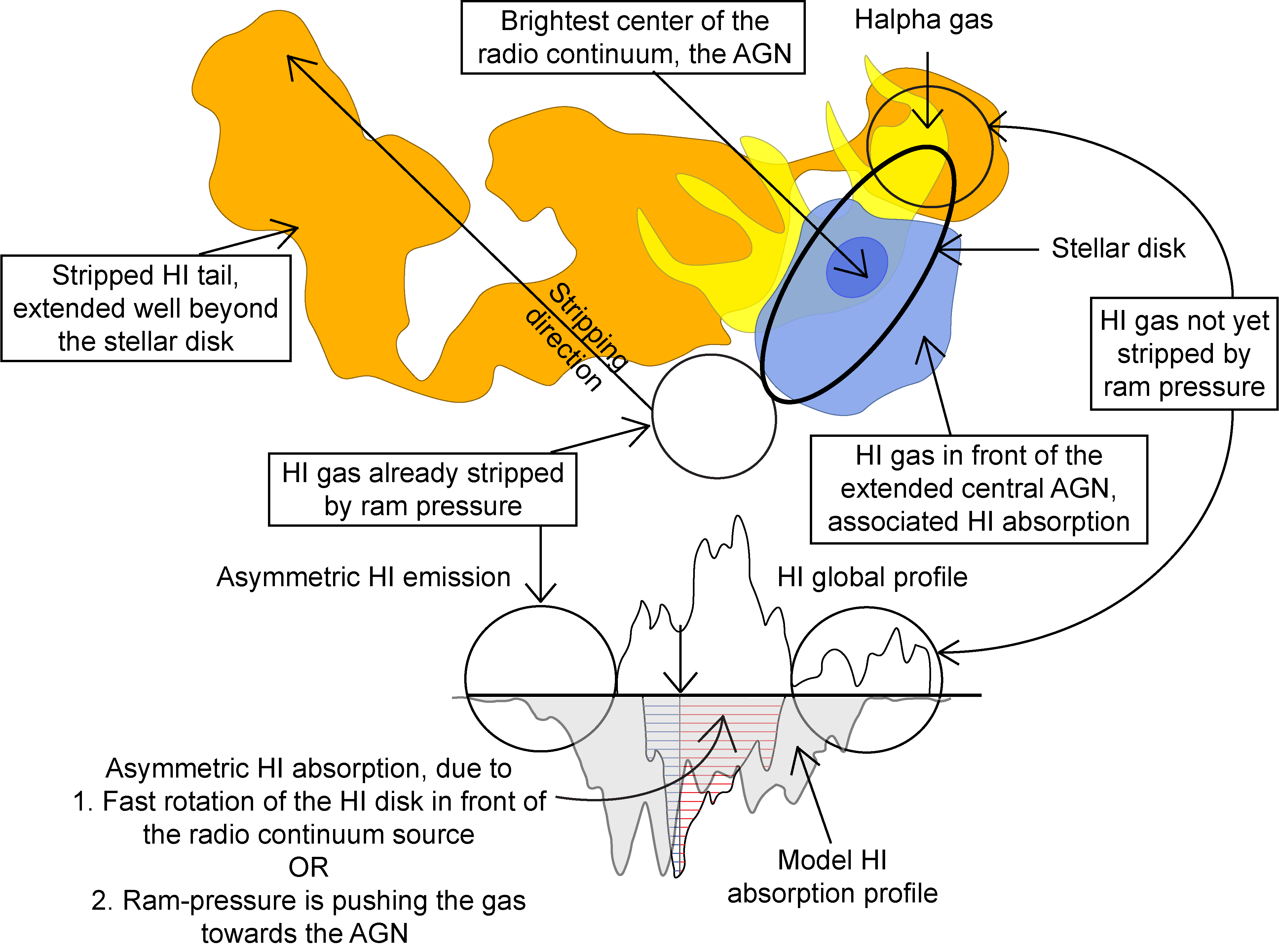}
    \caption{A schematic overview of the gas dynamics and ram-pressure stripping in JO204. Various aspects are described in the text.}
    \label{summary}
    \end{figure*}

\section{Discussion}

\par In a previous paper by \cite{Gullieuszik2017}, a wealth of multi-wavelength observations and a set of hydrodynamical simulations of JO204 were presented. Here we will briefly discuss how
the \HI\ observations fit in this picture and address some questions raised by our data.

\par Considering the scenario that might explain the redshifted wing of the observed \HI\ absorption profile, we conclude that the observed redshifted wing can be accommodated within the velocity range of the model absorption profile that is derived from the ionized gas kinematics seen in front of the central continuum source. If, in Fig \ref{abs_prof_HB_model_HI}, the grey profile of the modelled \HI\ absorption would have been narrower than the observed black curve then we could conclude that the additional observed \HI\ gas seen in absorption at higher velocities is pushed towards the radio continuum source by ram-pressure, triggering AGN activity. Since this is not the case, the observed redshifted \HI\ gas seen in absorption can not be unambiguously identified with cold gas being pushed towards the AGN. Observations with the JVLA in its A configuration could provide adequate angular resolution to resolve the continuum source and further test the hypothesis of ram-pressure induced AGN activity.

\par It is quite striking that the HI tail is stretching further out and in an apparently different direction compared to the H${\alpha}$ tail. We believe this east-ward extended region of the \HI\ tail was once in the southern part of the \HI\ disk of the galaxy. The direction of the H${\alpha}$ tentacles is similar to the north-eastern direction in the eastern-most part of the \HI\ tail (see also the stripping direction in Fig \ref{summary}). Furthermore, the \HI\ gas in the outer disk might also have been flung out radially after the gas has been pushed out of the midplane of the galaxy by ram-pressure.  Consequently, the centripetal gravitational force is weakened, causing an unbalanced centrifugal force. In this picture, we suggest that the stripped neutral gas from the southern part of the gas disk has formed the extended tail to the east of JO204 (Fig \ref{HI_map_of_JO204}).

\par Interestingly, from the BPT diagram (Fig 9 in \citealt{Gullieuszik2017}) we infer that the already stripped southern part of the disk may be exposed by the AGN ionization cone. We suggest that in JO204, the ablation of \HI\ gas clumps by the AGN ionization cone dissolves the gas clumps into diffuse ISM and the diffuse \HI\ gas can be more easily displaced by ram-pressure. Hence, we postulate that the AGN energy injection into the ISM is increasing the efficiency of gas loss through RPS and helping to create the extended tail in JO204.

\par The \HI\ line width in emission is narrower than the width as expected from the TF relation. But interestingly enough, the channel maps comparing the \HI\ and H$\alpha$ phases (Fig \ref{HI_on_Ha_JO204}) indicate that most of the \HI\ gas is out of the disk. Moreover, in the channels where we detect \HI\ emission in the north-western part of the disk, the velocities are higher than expected from the rotation (Fig \ref{global_profile}). Therefore, the gas may be spatially close to the midplane of the disk but it is kinematically decoupled from circular motion. Ram-pressure probably accelerates the gas in its circular orbit around the nucleus (like a ``tail wind"), thus the gas is pushed to higher velocities. So, we conclude that all of the \HI\ gas seen in emission, is decoupled from regular circular motion in the disk, kinematically or spatially. Certainly, the \HI\ mass of (1.32 $ \pm 0.13) \times 10^{9} \rm M_{\odot}$ that we calculated from our observation, is a lower limit on the total \HI\ mass of JO204 as the \HI\ gas in the disk is observationally `obscured' by the \HI\ absorption.

\par Fig \ref{HI_PV} shows that the \HI\ absorption covers a smaller velocity range around the systemic velocity than expected from the TF relation. This may happen when \HI\ gas is only present in the outer region of the disk, thus covering a smaller velocity range in front of the radio continuum source along our line of sight. It could mean that the ionized and \HI\ gas are not co-spatial since we know from Fig \ref{HB_PV} that the ionized gas places the galaxy right on the TF relation. However, to see the \HI\ gas from the outer disk in absorption, the galaxy should be edge on while JO204 is not an edge on galaxy. Therefore, as mentioned in Section 4.5, we believe that the \HI\ gas is very clumpy in the center of JO204. We also point out that the \HI\ absorption strength depends only on the strength of the radio continuum source and the optical depth of the medium. Clumpier \HI\ gas is denser, hence more easily detectable in absorption, while for the detection of \HI\ emission we are limited by the intensity of the emission and our detection threshold. Furthermore, we do not expect the inner disk to be affected by RPS because in that case, also the ionized gas disk would have been removed, which is clearly not the case. 

\par The 3$\sigma$ peak of  \HI\ emission in the north-western part of the disk recedes at a velocity that is larger than the TF-expected rotation velocity. As mentioned in a previous paragraph, the \HI\ gas from the outer spiral arms is probably stripped by ram-pressure, removed from the galactic disk potential well. This will result in a radial motion of the \HI\ gas due to a decreased centripetal force. Thus, this high velocity \HI\ gas seen in emission in the north-western side of the galaxy may be the signature of an unwinding spiral arm as a result of ram-pressure interaction with the ICM during cluster infall (Bellhouse et al. in prep).

\par Although the simulations by \cite{Gullieuszik2017} reproduce the H${\alpha}$ tail quite well, there is much more extraplanar \HI\ gas than H${\alpha}$ gas. The total \HI\ mass in the tail is (1.32 $ \pm 0.13) \times \rm 10^{9} \rm M_{\odot}$. It is really surprising that there is no \HI\ associated with many of the H${\alpha}$ knots. To explain the lack of coherence between these gas phases as evidenced by the channel maps, we hypothesize that the gas is probably highly turbulent in the tail and inter-mixed with new young stars that locally ionize the \HI\ gas to produce the H${\alpha}$ emission. The tail is a very chaotic medium with shocks, compression, star formation, ionization and diffusion into the ICM. It is not clear yet how star formation can happen in the tail and why it dies down farther away from the galaxy, where the turbulence may become too high, or perhaps the gas becomes too hot. We note, however, that the \HI\ gas has a low column density of $\rm 10^{20} \rm cm^{-2}$ at larger distances from the galaxy where there is no H${\alpha}$ emission. Hence, the gas density probably becomes too low to form stars that ionize the neutral gas. Thus, the \HI\ tail is further extended to the east in comparison with the H${\alpha}$ tail. 

\par The simulations by \cite{tonnesen2010,tonnesen2012} show that if gas condenses out of the tail producing star forming clumps, the peaks of \HI\ and H${\alpha}$ always coincide. Interestingly, from the preliminary analysis of ALMA observations of JO204 (Moretti et al. in prep), some CO is associated with the brightest H${\alpha}$ gas clumps. However, some H${\alpha}$ emission seems to occur without a CO counterpart. This suggests an efficient ablation of cold gas clumps in the tail of JO204. Overall, the physical mechanism responsible for non-coherence of the different gas phases will become clear by observing more jellyfish galaxies and comparing statistically the observational results with detailed hydrodynamical simulations. The situation varies from galaxy to galaxy depending on the geometry of the stripping episode and the physical state of both ISM and ICM, including any magnetic field. In JO206 (\citealt{ramatsoku2019}) there is a better match between the two gas phases. We are currently investigating the role played by magnetic fields in keeping the gas tail together (Mueller et al. in prep).

\par The stellar population analysis in \cite{Gullieuszik2017} shows that the stripping occurred sometime during the last 500 Myrs. We cannot obtain a more precise time estimate from the SINOPSIS analysis \citep{fritz2017}. This is what the top right panel of Fig 13 in \cite{Gullieuszik2017} shows: the H${\alpha}$ tail appeared between 6 $\times 10^{8}$ and 2$\times 10^{7}$ yrs ago. \cite{Gullieuszik2017}  also concluded that the galaxy was stripped from the outside in, i.e., gas from the outer disk had been removed first. The \HI\ data adds to this scenario. The disk is apparently stripped from south-east to north-west, the \HI\ has already been removed from the south-eastern part of the disk but is still intact in the north-western part. The stripped \HI\ is still visible as a tail to the east of the ionized gas. A rough estimate when the stripping of the \HI\ tail may have started is at least 360 Myrs ago assuming a velocity of 250 $\rm km s^{-1}$ (based on Fig \ref{HB_PV} and the TF relation) and maximum extent of the tail of 90 kpc.

\section{Summary}

\par In this work we have presented JVLA \HI\ observation of the jellyfish galaxy JO204, discovered from the ESO Large Programme GASP -- a survey of likely gas stripping galaxies. This work is focused on \HI\ observation of the JO204 galaxy and presents a comparison between the ionized \citep{Gullieuszik2017} and \HI\ gas kinematics.
 \begin{itemize}
 \item
 From the JVLA data, an \HI\ tail is observed that extends eastward beyond the stellar disk up to 90 kpc. The neutral gas tail is much longer than the ionized gas tail, which is 30 kpc based on MUSE observations. 
 
 \item
 When the neutral and ionized gas phases are compared across frequency channels, we observe that for some frequency channels, there is no neutral gas counterpart to the ionized gas, while in some cases, the neutral gas is offset from the ionized gas. For some frequency channels the neutral and ionized gas are coincident. This indicates that the gas is highly turbulent in the ram-pressure stripped tail, where scattered the young stars ionize the \HI\ gas locally to produce H$\alpha$ emission. 
 
 \item
 Considering the asymmetric \HI\ emission profile and the channel maps, we realize that there is still some neutral gas left in the northern-western part of the gas disk of JO204 while the neutral gas from the south-eastern part of the gas disk is already completely removed by ram-pressure.
 
 \item
 The asymmetric \HI\ absorption profile might be the \HI\ gas disk rotating at high velocity, coming in the line-of-sight of the small yet extended continuum source or ram-pressure is pushing the gas towards the AGN. The model neutral hydrogen absorption profile is produced assuming the same kinematics of ionized and \HI\ gas and using a Gaussian continuum source model from FIRST observation. The result shows larger velocity width of the model absorption profile in comparison with the observed \HI\ absorption. Since the column-density of the \HI\ gas across the patch through which the radio continuum source is observed is unknown, an approximation of a uniform column-density might have resulted this large velocity width. The \HI\ gas might be very clumpy covering only certain velocities, resulting a smaller width in velocity of the absorption profile.
 
 \item
  A schematic overview of the gas morphology and kinematics in JO204 is illustrated in Fig \ref{summary}. The light yellow and orange regions represent the ionized and neutral gas phases respectively. The small blue dot in the center indicates the brightest region of the radio loud AGN against which we see the \HI\ absorption, which is indicated in light blue and enlarged because of the JVLA-C beam size. The black ellipse outlines the stellar disk. The \HI\ tail extends well beyond the ionized gas tail. The north-western part of the gas disk still retains some neutral and ionized gas while the south-eastern part is already completely removed due to ram-pressure, creating an extended \HI\ tail in eastern direction and causing the asymmetry in the \HI\ emission profile. The \HI\ absorption has a redshifted wing which is modelled using ionized gas as representative of the \HI\ gas kinematics. The model \HI\ absorption profile is wider than the observed profile, indicating a clumpy nature of the \HI\ gas in the center.

\end{itemize}

\section*{Acknowledgements}
We would like to express our gratitude to the anonymous referee for helping us to improve the manuscript significantly. We would like to thank R. Morganti for her invaluable input about the \HI\ absorption modelling. T.D. would like to thank P. Bilimogga for useful discussions. This project has received funding from the European Research Council (ERC) under the European Union's Horizon 2020 research and innovation programme (grant agreement No. 833824). We acknowledge support by the Netherlands Foundation for Scientific Research (NWO) through VICI grant 016.130.338. We acknowledge financial support from PRIN-SKA 2017 (PI L. Hunt) and ``INAF main-streams" funding programme (PI B. Vulcani). This work is based on observations collected at the European Organisation for Astronomical Research in the southern hemisphere under ESO programme 196.B-0578 (PI B.M. Poggianti) and programme 0102.C-0589 (PI F. Vogt). Y.J. acknowledges financial support from CONICYT PAI (Concurso Nacional de Inserci\'on en la Academia 2017) No. 79170132 and FONDECYT Iniciaci\'on 2018 No. 11180558. This paper uses VLA data (Project code: VLA/17A-293), provided by the National Radio Astronomy Observatory which is a facility of the National Science Foundation operated under cooperative agreement by Associated Universities, Inc.

\bibliographystyle{mnras}
\bibliography{J0204}

%%%%%%%%%%%%%%%%%%%%%%%%%%%%%%%%%%%%%%%%%%%%%%%%%%



%\bibliographystyle{mnras}


%%%%%%%%%%%%%%%%%%%%%%%%%%%%%%%%%%%%%%%%%%%%%%%%%%

%%%%%%%%%%%%%%%%% APPENDICES %%%%%%%%%%%%%%%%%%%%%

%%%%%%%%%%%%%%%%%%%%%%%%%%%%%%%%%%%%%%%%%%%%%%%%%%

% Don't change these lines
\bsp	% typesetting comment
\label{lastpage}

\end{document}